\documentclass[pra,aps,amsmath,amssymb,showkeys]{revtex4}
\usepackage{graphicx}

\newcommand{\cli}{{\mathcal{I}}}

\newcommand{\mtm}{{\mathsf{L}}}
\newcommand{\mse}{{\mathsf{E}}}
\newcommand{\msg}{{\mathsf{G}}}
\newcommand{\msh}{{\mathsf{H}}}
\newcommand{\msj}{{\mathsf{J}}}
\newcommand{\msk}{{\mathsf{K}}}
\newcommand{\mtx}{{\mathsf{X}}}

\newcommand{\bro}{\boldsymbol{\rho}}
\newcommand{\vbro}{\boldsymbol{\varrho}}
\newcommand{\bmg}{\boldsymbol{\omega}}
\newcommand{\bsg}{\boldsymbol{\sigma}}
\newcommand{\bdl}{\boldsymbol{\delta}}
\newcommand{\Upn}{\Upsilon}
\newcommand{\vcr}{{\mathbf{r}}}
\newcommand{\vcs}{{\mathbf{s}}}
\newcommand{\pen}{\openone}
\newcommand{\tr}{\mathrm{tr}}
\newcommand{\dig}{\mathrm{diag}}

\newcommand{\suc}{\mathrm{suc}}

\newcommand{\ron}{{\mathrm{ran}}}
\newcommand{\clm}{{\mathcal{M}}}
\newcommand{\cpt}{{\mathtt{C}}}
\newcommand{\iu}{{\mathtt{i}}}
\newcommand{\wamma}{\bar{\gamma}}
\newcommand{\xdif}{{\mathrm{d}}}

\begin{document}
\clearpage
\preprint{}

\title{Quantum search degeneration under amplitude noise in queries to the oracle}

\author{Alexey E. Rastegin, Anzhelika M. Shemet}

\affiliation{Department of Theoretical Physics, Irkutsk State University, Irkutsk 664003, Russia}

\begin{abstract}
We examine how amplitude noise in queries to the oracle degrades a
performance of quantum search algorithm. The Grover search and
similar techniques are widely used in various quantum algorithms,
including cases where rival parties are fighting over confidential
data. Hence, the oracle-box wires become the subject of competing
activity with an alteration of their normal functioning. Of course,
many kinds of errors could arise in this way. Possible influence of
dephasing on quantum search was already addressed in the literature.
Amplitude damping is another type of errors that should be analyzed
first. To study the problem, we introduce a simple model of
collective distortions with the use of amplitude damping channel.
All the quantities of interest should be considered as functions of
the number of Grover's iterations. In particular, we investigate the
success probability with respect to the parameter that characterizes
the level of amplitude errors. The success probability degrades
significantly even if the error amount is not essential. Namely,
this probability soon enough reduces to a value close to one half.
We also study trade-off relations between quantum coherence and the
success probability in the presence of amplitude noise.
\end{abstract}


\maketitle

\pagenumbering{arabic}
\setcounter{page}{1}

\section{Introduction}\label{sec1}

Today, the use of quanta as a tool for information processing has
found a great attention \cite{wendin17,flamini19,sage19,henriet20}.
Although quantum technologies are still in the status of emerging, a
stable progress is observed in both theory and practice. Celebrated
Shor's result \cite{shor97} gave a stimulus to numerous algorithms
for algebraic problems \cite{vandam2010}. Grover's search algorithm
\cite{grover97,grover97a,grover98} is another fundamental result in
quantum information science. Now, the technique of amplitude
amplification is one of cornerstones in building quantum algorithms
\cite{childsQA}. The Grover algorithm is optimal for search by means
of queries to the oracle \cite{bbbv97,zalka99}. That is, we invoke
the oracle to process any item, whereas the database itself is not
exposed explicitly. The original formulation of Grover has later
been modified with more general blocks and an arbitrary initial
distribution. Various generalizations of quantum search were
considered in the literature
\cite{biham99,mardel00,biham2002,apati2016,suzuki21}.

By the oracle, we mean some black box able to calculate values of
the desired Boolean function. Any user queries the box by putting
concrete values of the argument. In practice, an access to the
oracle cannot be treated as exceptional and even reliable. Rather, a
number of simultaneous queries will be expected here. Moreover,
users' goals may be different and ever opposite. These are reasons
for which errors in the oracle-box wires should be studied. There
are various possible scenarios to analyze the above questions. Using
simplified model of collective amplitude errors, we address one of
such scenarios. In a certain sense, the treatment of \cite{rast18}
is reformulated with another kind of noise. Another approach to
quantum search under localized dephasing was considered in
\cite{reitzner19}. We also examine the relative entropy of coherence
from the viewpoint of its trade-off with the success probability.

This paper is organized in the following way. The preliminary facts
are discussed in Sect. \ref{sec2}. In Sect. \ref{sec3}, we describe
and motivate the model of amplitude errors that occur in the
oracle-box wires. The built model leads to the recursion equation in
terms of the effective Bloch vector. In Sect. \ref{sec4}, we examine
changes of the success probability after repeated Grover's
iterations under amplitude noise. It is demonstrated through
visualization that the Grover search algorithm is sensitive to
errors of the considered type. Using the relative entropy of
coherence, we investigate trade-off relations of register coherence
with the success probability. In Sect. \ref{sec5}, we conclude the
paper with a summary of the results. Appendices are devoted
to derivation of auxiliary formulas.

\section{Preliminaries}\label{sec2}

Let us begin with the original formulation of Grover's search
algorithm. It can be posed as follows. The search space contains
$N=2^{n}$ binary strings $x=(x_{1}\cdots{x}_{n})$ with
$x_{j}\in\{0,1\}$ so that $x\in\{0,1,\ldots,N-1\}$. The problem is
to find one of marked strings that form some set $\clm$, whereas
other strings are all in the complement $\clm^{\cpt}$. With no loss
of generality, one can assume $1\leq|\clm|\leq{N}/2$. To check the
given string $x$, the algorithm appeals to the oracle which returns
the value of Boolean function $x\mapsto{F}(x)$ such that $F(x)=1$
for $x\in\clm$ and $F(x)=0$ for $x\in\clm^{\cpt}$. The algorithm
initializes the $n$-qubit register to $|0\rangle$ and then apply the
Hadamard transform, whence the output
\begin{equation}
\msh|0\rangle=\frac{1}{\sqrt{N}}\sum_{x=0}^{N-1} |x\rangle
\, . \label{uniam}
\end{equation}
Superpositions of such a kind are necessary to realize the quantum
parallelism \cite{deutsch85}. Further, we repeat the Grover
iteration involving two steps. The first step with querying the
oracle can be represented by the rotation operator
\begin{equation}
\msj=\sum_{x=0}^{N-1} (-1)^{F(x)}\,|x\rangle\langle{x}|
\, . \label{jfpi}
\end{equation}
Due to (\ref{jfpi}), the amplitudes of marked states are multiplied
by the factor $\exp(\iu\pi)=-1$. The second step of the Grover
iteration leads to the inversion about mean \cite{nielsen}. It is
represented by the operator
\begin{equation}
\msk=2\msh|0\rangle\langle0|\msh-\pen_{N}
\, , \label{roj0}
\end{equation}
where $\pen_{N}$ is the identity operator of the corresponding size.
So, the standard Grover iteration is expressed as:
\begin{equation}
\msg=\msk\msj
\, . \label{msg0}
\end{equation}
In the standard formulation, the initial distribution of amplitudes
is taken in the form (\ref{uniam}). Then the evolution of amplitudes
is described within the two-dimensional picture, for which we use
the normalized superpositions of unmarked and marked states, viz.
\begin{align}
|w\rangle&:=\frac{1}{\sqrt{N-M}}\sum_{x\in\clm^{\cpt}} |x\rangle
\, , \label{sunm}\\
|m\rangle&:=\frac{1}{\sqrt{M}}\sum_{x\in\clm} |x\rangle
\, . \label{smar}
\end{align}
It is useful to put the parameter $\theta\in(0,\pi/2)$ such that
$\cos\theta=1-2M/N$ and
\begin{equation}
\sin^{2}\theta/2=\frac{M}{N}
\ , \qquad
\cos^{2}\theta/2=1-\frac{M}{N}
\ . \label{sicos0}
\end{equation}
Writing operators as matrices in the basis
$\bigl\{|w\rangle,|m\rangle\bigr\}$ results in
$\msj=\dig(+1,-1)=\bsg_{z}$ and
\begin{equation}
\msk=
\begin{pmatrix}
    \cos\theta & \sin\theta \\
    \sin\theta & -\cos\theta
\end{pmatrix}
 . \label{kmat}
\end{equation}
The operator (\ref{msg0}) is therefore represented as
\cite{nielsen}:
\begin{equation}
\msg=
\begin{pmatrix}
    \cos\theta & -\sin\theta \\
    \sin\theta & \cos\theta
\end{pmatrix}
 . \label{grot0}
\end{equation}
The initialization state (\ref{uniam}) is represented as
$\cos(\theta/2)|w\rangle+\sin(\theta/2)|m\rangle$. It is very close
to $|w\rangle$ due to smallness of the angle $\theta$ in typical
situations. Without noise, each Grover iteration rotates the
register state by $\theta$ towards the superposition $|m\rangle$. To
characterize a behavior of quantum search under noise, we restrict
our consideration to a sufficiently simple model of errors. This
choice allows us to answer explicitly basic questions of the study.

Quantum speed-up seems to be impossible without entanglement
\cite{bpati2002,jozsa03}. To analyze quantum algorithms,
correlations in register states are considered with respect to few
prescribed bases. Hence, the framework to quantify quantum coherence
as a potential resource should be developed  \cite{bcp14,plenio16}.
The corresponding resource theory allows one to use coherence as a
diagnostic tool for quantum chaos \cite{anand21}. In algorithms with
amplitude amplification, trade-off relations between quantum
coherence and the success probability is one of important questions.
We shall address this issue in the case, when queries to the oracle
are exposed to amplitude noise of the considered type. In general,
various approaches to measure quantum correlations were discussed in
the literature \cite{abc16,plenio16,hufan16}. Each candidate to
quantify the level of coherence is based on some measure to
distinguish quantum states. In this paper, the relative entropy of
coherence will be used as a quantitative characteristics.

Let us consider the set $\cli$ of all density matrices of the form:
\begin{equation}
\bdl=\sum_{x=0}^{N-1} b(x)\,|x\rangle\langle{x}|
\, , \qquad
\sum_{x=0}^{N-1} b(x)=1
\, . \label{incd}
\end{equation}
Such matrices are incoherent in the computational basis. It is
further asked how far the given state is from states of the set
$\cli$. The quantum relative entropy of $\bro$ with respect to
$\bmg$ is defined as \cite{vedral02}:
\begin{equation}
D_{1}(\bro||\bmg):=
\begin{cases}
\tr(\bro\ln\bro\,-\bro\ln\bmg) \,,
& \text{if $\ron(\bro)\subseteq\ron(\bmg)$} \, , \\
+\infty\, , & \text{otherwise} \, .
\end{cases}
\label{relan}
\end{equation}
By $\ron(\bro)$, one means the range of $\bro$. Using (\ref{relan}),
the relative entropy of coherence is introduced as \cite{bcp14}:
\begin{equation}
C_{1}(\bro):=
\underset{\bdl\in\cli}{\min}\,D_{1}(\bro||\bdl)
\, . \label{c1df}
\end{equation}
After minimization, we obtain the formula \cite{bcp14}
\begin{equation}
C_{1}(\bro)=S_{1}(\bro_{\dig})-S_{1}(\bro)
\, , \label{c1for}
\end{equation}
where $S_{1}(\bro)=-\,\tr(\bro\ln\bro)$ is the von Neumann entropy
of $\bro$. The closest incoherent state is expressed as:
\begin{equation}
\bro_{\dig}:=\sum_{x=0}^{N-1} p(x)\,|x\rangle\langle{x}|
\, ,
\qquad
p(x)=\langle{x}|\bro|x\rangle
\, . \nonumber
\end{equation}
For basic properties of (\ref{c1df}), see the papers
\cite{bcp14,plenio16}. Generalized entropic functions are widely
used in quantum information science. For this reason, we mark the
above quantities by the subscript $1$. Coherence quantifiers induced
by quantum divergences of the Tsallis type were examined in
\cite{rastpra16}. Coherence monotones based on R\'{e}nyi divergences
were considered in \cite{chitam2016,shao16,skwgb16}. Other
candidates to quantify the amount of coherence were examined in
\cite{shao15,rpl15}. The geometric coherence is an interesting
quantifier of different kind \cite{plenio16}. Complementarity
relations for quantum coherence were formulated in several ways
\cite{hall15,pati16,baietal6,rastcomu}, including duality between
the coherence and path information
\cite{bera15,bagan16,bagan20,qureshi21}. From the computational
viewpoint, the concept of quantum coherence was examined in
\cite{hillery16,hfan2016}. In particular, the authors of
\cite{hfan2016} reported on coherence depletion in the original
Grover algorithm. Relations between coherence and the success
probability in generalized amplitude amplification were studied in
\cite{rastgro17}. Some of these results will be used to analyze
Grover's search in the presence of amplitude errors.

\section{Collective errors introduced by amplitude damping}\label{sec3}

There exist many scenarios of decoherence of quantum computations,
even if only the oracle-box wires are exposed to noise. Let us
consider a model with amplitude damping. Following \cite{rast18}, we
restrict a consideration to density matrices effectively
two-dimensional with respect to the basis
$\bigl\{|w\rangle,|m\rangle\bigr\}$. Such matrices can be
represented via the Bloch vector $\vcr=(r_{x},r_{y},r_{z})$, so that
\begin{equation}
\bro=\frac{1}{2}
\begin{pmatrix}
    1+r_{z} & r_{x}-\iu{r}_{y} \\
    r_{x}+\iu{r}_{y} & 1-r_{z}
\end{pmatrix}
 . \label{dmatr}
\end{equation}
With positive parameter $\gamma\leq1$, we introduce the following
Kraus operators:
\begin{equation}
\mse_{0}:=
\begin{pmatrix}
    1 & 0 \\
    0 & \sqrt{1-\gamma}
\end{pmatrix}
 , \qquad
\mse_{1}:=
\begin{pmatrix}
    0 & \sqrt{\gamma} \\
    0 & 0
\end{pmatrix}
 . \label{kraus01}
\end{equation}
These operators describe the action of amplitude damping
$\Phi_{\mse}$ on density matrices of the considered type. It is easy
to see that \cite{nielsen}
\begin{equation}
(r_{x},r_{y},r_{z})\overset{\Phi_{\mse}\,}\longmapsto
\bigl(\sqrt{1-\gamma}\,r_{x},\sqrt{1-\gamma}\,r_{y},\gamma+(1-\gamma)r_{z}\bigr)
\, . \label{blov}
\end{equation}
Initializing leads to the density matrix
$\bro(0)=\msh|0\rangle\langle0|\msh$ with the Bloch vector
$\vcr(0)=(\sin\theta,0,\cos\theta)^{\mathsf{T}}$. After $t$
iterations, the success probability is expressed as:
\begin{equation}
P_{\suc}(t)=\langle{m}|\bro(t)|m\rangle=\frac{1-r_{z}(t)}{2}
\ . \label{psuct}
\end{equation}
The following fact is seen from (\ref{blov}) and (\ref{psuct}). When
$\gamma>0$, the map $\Phi_{\mse}$ does not alter the success
probability only for $r_{z}=1$. The latter corresponds to the
superposition $|w\rangle$ of unmarked states.

Let us proceed to the case of noise in the oracle-box wires
\cite{rast18}. Each iteration acts on density matrices of the
register according to the formula:
\begin{equation}
\bro(t)\mapsto\bro(t+1)
=\Upn_{\msk}\circ\Phi_{\mse}\circ\Upn_{\msj}\circ\Phi_{\mse}\bigl(\bro(t)\bigr)
\, . \label{iteralt}
\end{equation}
The two unitary channels read here as
$\Upn_{\msj}(\vbro)=\msj\,\vbro\,\msj^{\dagger}$ and
$\Upn_{\msk}(\vbro)=\msk\,\vbro\,\msk^{\dagger}$. For $\gamma=0$,
the map $\Phi_{\mse}$ acts as identical, whence the iteration
becomes
\begin{equation}
\bro(t+1)=\Upn_{\msk}\circ\Upn_{\msj}\bigl(\bro(t)\bigr)
=\msg\,\bro(t)\,\msg^{\dagger}
\, . \label{iterc}
\end{equation}
Applying consequently the map (\ref{iterc}) to the initial state
(\ref{uniam}), we will always deal with pure states of the register.
It is not the case for the altered map (\ref{iteralt}). The above
model deals with collective amplitude distortions expressed in the
computational basis. This approach allows us to formulate results in
a closed analytic form. Also, our scenario does not mean that
distortions are similar for all qubits \cite{rast18}.

The following method will be used to solve (\ref{iteralt}). Since
only two components of the Bloch vector are non-zero, we further
treat $\vcr(t)$ as a column with two entries
\begin{equation}
\vcr(t)=
\begin{pmatrix}
    r_{x}(t) \\
    r_{z}(t)
\end{pmatrix}
 . \label{vcr2}
\end{equation}
Then the action of operation $\Phi_{\mse}$ is represented by the
formula
\begin{equation}
\vcr\overset{\Phi_{\mse}\,}\longmapsto
\dig\bigl(\sqrt{1-\gamma},1-\gamma\bigr)\,\vcr+\gamma\,(0\ 1)^{\mathsf{T}}
\, . \label{pseb}
\end{equation}
We also write
\begin{equation}
2\,\msj\,\bro\,\msj^{\dagger}=\bsg_{z}(\pen_{2}+r_{x}\bsg_{x}+r_{z}\bsg_{z})\bsg_{z}
=\pen_{2}-r_{x}\bsg_{x}+r_{z}\bsg_{z}
\, , \label{bsxz}
\end{equation}
whence the operation $\Upn_{\msj}$ acts on Bloch vectors as the
matrix $\dig(-1,+1)$. Due to (\ref{kmat}), one has
$\msk\,\pen_{2}\,\msk^{\dagger}=\pen_{2}$,
\begin{align}
\msk\,\bsg_{x}\msk^{\dagger}&=-\cos2\theta\,\bsg_{x}+\sin2\theta\,\bsg_{z}
\, , \nonumber\\
\msk\,\bsg_{z}\msk^{\dagger}&=\sin2\theta\,\bsg_{x}+\cos2\theta\,\bsg_{z}
\, , \nonumber
\end{align}
so that $r_{x}\mapsto-\cos2\theta\,r_{x}+\sin2\theta\,r_{z}$ and
$r_{z}\mapsto\sin2\theta\,r_{x}+\cos2\theta\,r_{z}$ under the map
$\Upn_{\msk}$. Hence, this operation acts on Bloch vectors as the
matrix
\begin{equation}
\begin{pmatrix}
    -\cos2\theta & \sin2\theta \\
    \sin2\theta & \cos2\theta
\end{pmatrix}
 . \label{kvbk}
\end{equation}
Finally, the recursion equation is written in terms of the
effective Bloch vector as:
\begin{equation}
\vcr(t+1)=\wamma\,\mtm\,\vcr(t)+(1-\wamma^{2})
\begin{pmatrix}
    \sin2\theta \\
    \cos2\theta
\end{pmatrix}
 , \label{itrr}
\end{equation}
where $\wamma=1-\gamma$ and the matrix $\mtm$ reads as:
\begin{equation}
\mtm=
\begin{pmatrix}
    \cos2\theta & \wamma\sin2\theta \\
    -\sin2\theta & \wamma\cos2\theta
\end{pmatrix}
 . \label{matra}
\end{equation}
We further decompose the Bloch vector as:
\begin{equation}
\vcr(t)=\vcs(t)+(1-\wamma^{2})\bigl(\pen_{2}-\wamma\,\mtm\bigr)^{-1}\!
\begin{pmatrix}
    \sin2\theta \\
    \cos2\theta
\end{pmatrix}
 . \label{ritr}
\end{equation}
It is possible whenever
\begin{equation}
\det\bigl(\pen_{2}-\wamma\,\mtm\bigr)=1-(\wamma+\wamma^{2})\cos2\theta+\wamma^{3}\neq0
\, . \label{detg}
\end{equation}
It follows from Appendix \ref{app0} that this determinant is
strictly positive for $\cos2\theta<1$ and all $\wamma\in[0,1]$. Then
the equation (\ref{itrr}) leads to
\begin{equation}
\vcs(t+1)=\wamma\,\mtm\,\vcs(t)
\, , \qquad
\vcs(t)=\wamma^{t}\,\mtm^{t}\vcs(0)
\, . \label{vcsv}
\end{equation}
For $\gamma>0$, the vector $\vcs(t)$ becomes zero in the limit
$t\to\infty$. Then the Bloch vector coincides with the second term
in the right-hand side of (\ref{ritr}). This column is explicitly
given in (\ref{vcrt}), whence
\begin{equation}
\lim_{t\to\infty}P_{\suc}(t)
=\frac{1}{2}+\frac{(1-\wamma^{2})(\wamma-\cos2\theta)}{2\bigl(1-(\wamma+\wamma^{2})\cos2\theta+\wamma^{3}\bigr)}
\ . \label{psucin}
\end{equation}
The latter depends on all the parameters $\gamma$, $M$ and $N$. In
Appendix \ref{app1}, the components $s_{x}(t)$ and $s_{z}(t)$ are
represented in detail. In this way, we express quantities of
interest as functions of iterations.

Let us discuss briefly a relevance of the considered noise model to
real hardware gates. There are several platforms to obtain quantum
processors in the noisy intermediate-scale quantum (NISQ) era
\cite{preskill18}, e.g., superconducting qubits \cite{wendin17},
trapped ions \cite{sage19}, neutral atoms \cite{henriet20}. Overall,
for single-qubit operations high efficiencies can be reached within
all these technologies. However, the question of characterization of
NISQ devices is very complicated due to a wide range of
possibilities for sizing, connection topology and gate
implementations. Recent analysis of trapped-ion hardware reported
the following \cite{debroy20}. With growing the number of qubits, a
gate fidelity crucially depends on the type of performed algorithms.
On average, they are of the same order or even lesser than will be
used in examples of the next section. The noise model posed by
(\ref{itrr}) is effectively two-dimensional, but the qubit register
as a whole undergoes collective errors. It turns out that quite low
level of errors is required to implement the search process
efficiently. The results of \cite{debroy20} give a reason in favor
of actual partition of register states into blocks with very
different dealing. The role of such blocks in quantum search with
noise was already emphasized in \cite{reitzner19}.

\section{On dynamics of the success probability and quantum coherence}\label{sec4}

In this section, we will use the above model to study Grover's
search under amplitude noise in the oracle-box wires. It is
instructive to visualize $P_{\suc}(t)$ as a function of $t$ for
several values of the parameters $\gamma$ and $N$. Before that, we
shortly discuss some general features of search under amplitude
noise of the considered type. In general, the character of
functional behavior depends on relations between the actual values
of $\gamma$, $M$ and $N$. The formulas (\ref{sxtfin1}) and
(\ref{sztfin1}) express the components of $\vcs(t)$ via
trigonometric functions under the condition (\ref{imdi}). The
quantities of interest show an oscillating behavior with slowly
decreasing amplitudes. We will also assume the most typical
situation $M\ll{N}$, whence $\theta\approx2\sqrt{M/N}$ due to
(\ref{sicos0}). It is advisable to focus on very small $\gamma$,
when the actual search process does not deviate essentially from the
idealized one. Moreover, a parametric dependence in this case
becomes clearer. For $\gamma=0$, each Grover iteration rotates
clockwise the Bloch vector by the angle $2\theta$. It follows from
(\ref{ph2th}) that $\varphi=2\theta+O(\gamma^{2})$. For sufficiently
small $\gamma$, one can write
\begin{align}
s_{x}(t)&\simeq\wamma^{3t/2}\bigl[\sin(2t+1)\theta+O(\gamma)\bigr]
\, , \label{sxtsim}\\
s_{z}(t)&\simeq\wamma^{3t/2}\bigl[\cos(2t+1)\theta+O(\gamma)\bigr]
\, . \label{sztsim}
\end{align}
We refrain from presenting the calculations here. Oscillations of
$P_{\suc}(t)$ around its effective mean are determined by
(\ref{sztsim}). What happens with this term, when small $\gamma$
grows at fixed $M$ and $N$? Here, the number of iterations to reach
the $k$-th peak of $P_{\suc}(t)$ is such that
\begin{equation}
t_{k}\theta\simeq\frac{(2k-1)\pi}{2}
\ , \label{t1df}
\end{equation}
whence $\cos(2t_{k}+1)\theta\simeq-\cos\theta$. The following fact
is seen from (\ref{sztsim}). At the given $t$, deviations of
$r_{z}(t)$ from its effective mean are mainly defined by
$(1-\gamma)^{3t/2}=\exp(-\alpha{t})$ with
$\alpha=-(3/2)\ln(1-\gamma)\approx3\gamma/2$. The number $t_{k}$
remains almost constant, whereas the ratio of deviation amplitudes
has a behavior
\begin{equation}
\frac{s_{z}(t_{k})}{s_{z}^{\prime}(t_{k})}\simeq\exp\bigl\{(\alpha^{\prime}-\alpha)t_{k}\bigr\}
\, . \label{a1a1p}
\end{equation}
Hence, variations of $P_{\suc}(t)$ are also suppressed with growth
of small $\gamma$. This decay is larger for further peaks. The
obtained conclusions seem to be quite natural.

The picture is more complicated, when $N$ increases and other
parameters are fixed. Due to (\ref{t1df}), we write
\begin{equation}
t_{k}\simeq\frac{(2k-1)\pi}{4}\,\sqrt{\frac{N}{M}}
\ . \label{t1df1}
\end{equation}
Thus, the number of iterations to reach the $k$-th peak increases
proportionally to the square of $N$. Instead of (\ref{a1a1p}), we
herewith obtain
\begin{equation}
\frac{s_{z}(t_{k})}{s_{z}^{\prime}(t_{k}^{\prime})}\simeq
\exp\!\left\{\frac{(2k-1)\pi\alpha}{4\sqrt{M}}\,\bigl(\sqrt{N^{\prime}}-\sqrt{N}\,\bigr)\right\}
 . \label{a1a1pn}
\end{equation}
For $N^{\prime}>N$, the spread of variations becomes lesser. In
other words, the curve of $P_{\suc}(t)$ is stretched with a
simultaneous reduce of deviations. Volumizing the search space
results in increasing the required iterations jointly with
decreasing the actual success probability. This character of changes
could be expected without analysis.

\begin{figure}
\includegraphics[height=7.2cm]{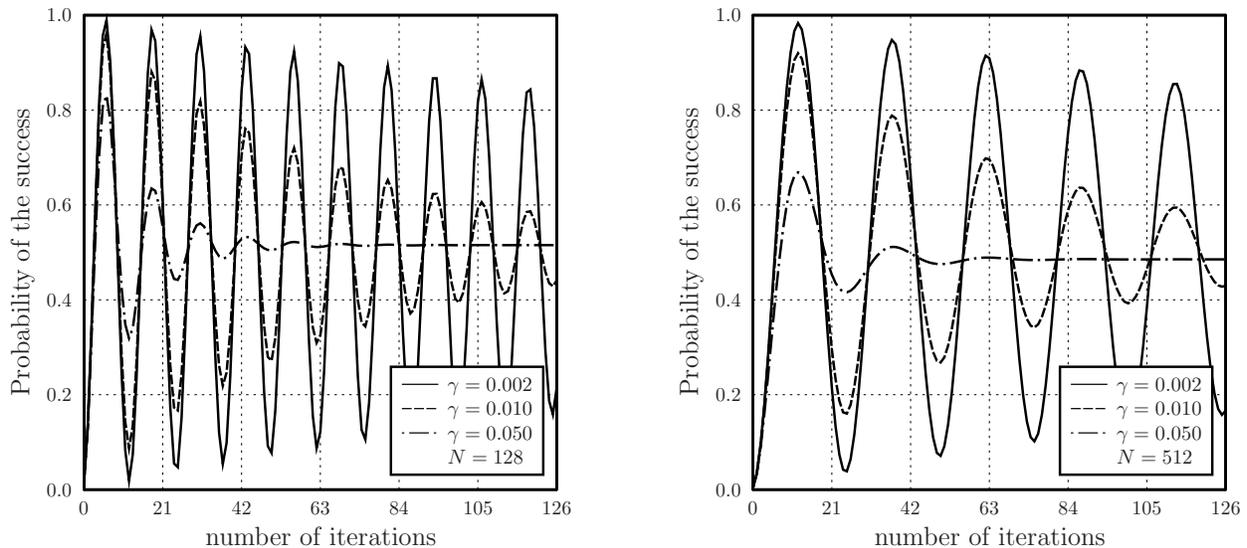}
\caption{\label{fig1} The function $P_{\suc}(t)$ for few values of
$\gamma$ and $M=2$, with $N=128$ on the left and $N=512$ on the
right.}
\end{figure}

To illustrate the above findings, the quantity $P_{\suc}(t)$ is
shown in Fig. \ref{fig1} for few small values of $\gamma$. The left
and right boxes respectively draw the lines for $N=128$ and $N=512$,
whereas $M=2$ in both the cases. In accordance with (\ref{t1df1}),
the increase of $N$ by four times  implies that the number of peaks
on the left is the doubled number of peaks on the right. In
addition, consecutive peaks of the success probability decay faster
in the right box. We also observe that the limit $t\to\infty$ leads
to different values. For fixed $M$, these limiting values depend on
both $\gamma$ and $N$ according to (\ref{psucin}). For very small
$\gamma$, a dependence on $N$ is revealed more brightly. In Fig.
\ref{fig1}, we restrict a consideration to $\gamma$ up to five
hundredths. Increasing $\gamma$ somewhat further, we have come
across a change of functional behavior from trigonometric to
hyperbolic type. The formulas (\ref{sxtfin1}) and (\ref{sztfin1})
are then replaced with (\ref{sxtfin11}) and (\ref{sztfin11}),
respectively. In this case, even the first peak of $P_{\suc}(t)$
turns out to be suppressed essentially. Under such circumstances,
the search process actually becomes of no effect. Thus, the
appearance of amplitude errors even only in the oracle-box wires can
lead to devaluation of the quantum search.

The role of quantum coherence in quantum search has found a lot of
attention in the literature. It was shown in \cite{rastgro17} that
\begin{equation}
h_{1}(P_{\suc})\leq{C}_{1}(\bro)+S_{1}(\bro)\leq
P_{\suc}\,\ln\!\left(\frac{M}{P_{\suc}}\right)+
(1-P_{\suc})\,\ln\!\left(\frac{N-M}{1-P_{\suc}}\right)
 , \label{renp02}
\end{equation}
where $h_{1}(P_{\suc})$ is the binary Shannon entropy. To calculate
$C_{1}\bigl(\bro(t)\bigr)$, we write the actual density matrix as:
\begin{equation}
\bro(t)=\frac{1+r_{z}(t)}{2}\> |w\rangle\langle{w}|
+\frac{r_{x}(t)}{2}\,\Bigl(|w\rangle\langle{m}|+|m\rangle\langle{w}|\Bigr)
+\frac{1-r_{z}(t)}{2}\> |m\rangle\langle{m}|
\, . \label{rhtwm}
\end{equation}
We ask for the diagonal part of $\bro(t)$ in the computation basis.
It is represented by the diagonal line, in which the value
$\bigl(1-P_{\suc}(t)\bigr)/(N-M)$ stands $(N-M)$ times and the value
$P_{\suc}(t)/M$ stands $M$ times. For $t>0$, the non-zero
eigenvalues of $\bro(t)$ read as:
\begin{equation}
\frac{1\pm\|\vcr(t)\|_{2}}{2}
\ , \qquad
\|\vcr(t)\|_{2}=\sqrt{r_{x}(t)^{2}+r_{z}(t)^{2}}\leq1
\, . \label{eigrt}
\end{equation}
With the initial distribution (\ref{uniam}) we have
$\|\vcr(0)\|_{2}=1$. In the considered case of amplitude noise, one
has
\begin{equation}
C_{1}\bigl(\bro(t)\bigr)=
P_{\suc}(t)\,\ln\!\left(\frac{M}{P_{\suc}(t)}\right)
+\bigl(1-P_{\suc}(t)\bigr)\ln\!\left(\frac{N-M}{1-P_{\suc}(t)}\right)-S_{1}\bigl(\bro(t)\bigr)
\, . \label{trad}
\end{equation}
Here, the upper bound of the right-hand side of (\ref{renp02}) is
saturated. For the given $P_{\suc}$, this upper bound provides the
maximal possible value of $C_{1}(\bro)$. We see from (\ref{trad}) a
trade-off between $C_{1}\bigl(\bro(t)\bigr)$ and $P_{\suc}(t)$. When
the success probability approaches $1$, the relative entropy of
coherence decreases. In the absence of noise, the relative entropy
of coherence fluctuates between the maximum and minimum values
roughly evaluated as $\ln(N-M)$ and $\ln{M}$. They occur when the
success probability is close to its extreme values.

\begin{figure}
\includegraphics[height=7.2cm]{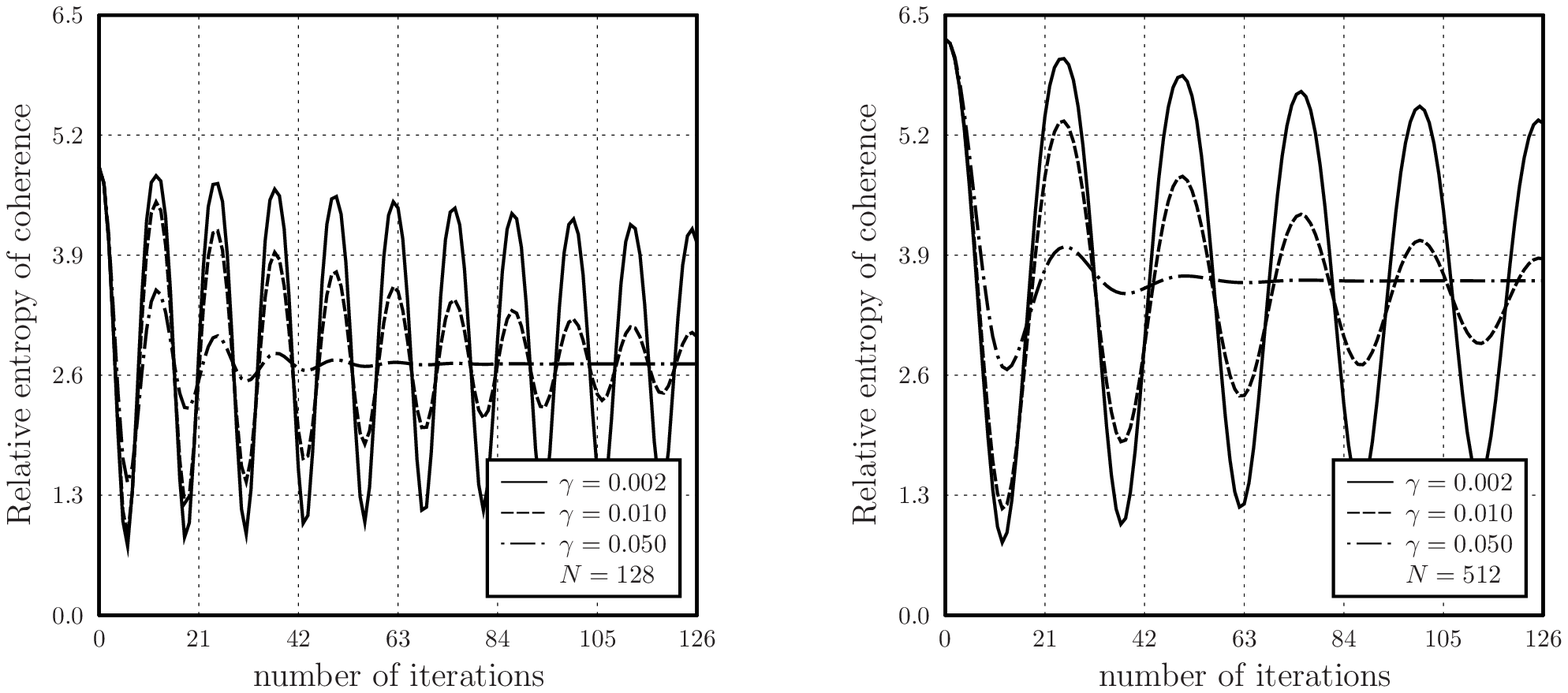}
\caption{\label{fig2} The function $C_{1}\bigl(\bro(t)\bigr)$ for
few values of $\gamma$ and $M=2$, with $N=128$ on the left and
$N=512$ on the right.}
\end{figure}

Amplitude errors lead to suppressing oscillations of all the
quantities including the relative entropy of coherence. Of course,
the corresponding decay is intensified with growth of $\gamma$. Let
us visualize a dependence of the relative entropy of coherence on
iterations. Figure \ref{fig2} shows $C_{1}\bigl(\bro(t)\bigr)$ under
the same circumstances as in Fig. \ref{fig1}. For convenience we
note $\ln(N-M)\approx4.836$ for $N=128$ and $\ln(N-M)\approx6.234$
for $N=512$. These numbers approximately characterize maximal values
for small $\gamma$ and such iterations that oscillation decay has
not yet had time to manifest. Minimal values will be close to
$\ln2\approx0.693$ for vanishing noise. Of course, we again observe
all the features mentioned in connection with Fig. \ref{fig1}. In
particular, the increase of $N$ by four times results in doubling
the period with respect to $t$. Further, consecutive peaks of the
relative entropy of coherence are attenuated faster in the right
box. Asymptotically, the relative entropy of coherence tends to fit
a constant. Then the search process becomes completely degenerated.

\section{Conclusions}\label{sec5}

We have examined the case, when queries to the oracle in Grover's
search algorithm are exposed to amplitude error of the specific
type. The used model of noise is inspired by the amplitude damping
channel reformulated with respect to the two subsets of the unmarked
and marked states, respectively. The presented results aim to extend
and complete the analysis initiated in the paper \cite{rast18}. That
paper is devoted to degradation of quantum search under the action
of collective phase flips in the oracle-box wires. More general
scenarios of quantum search under localized dephasing were examined
in \cite{reitzner19}. Despite of its simplicity, the used model
allows us to probe specific features of search degeneration under
amplitude noise in queries to the oracle. In general, amplitude
distortions are inevitable in real communication lines. Further,
characteristics of NISC devices are expected to be dependent on
the algorithm type \cite{debroy20}.

The following conclusions can be formulated on the base of the
considered model of amplitude noise and the results of
\cite{rast18}. Since the Grover algorithm needs very multiple
queries to the oracle, the quantum search process soon degenerates
even with relatively moderate noise levels in the oracle-box wires.
In addition to the type of arising noise, we mention the role of
localization of collective errors in the sense of dividing states of
the computational basis into different blocks. If there are concerns
about noise or opposite activity in the channel of link to the
oracle, then it is quite reasonable to limit iterations to a number
sufficient to get a vicinity of the first peak of the success
probability. Apparently, legitimate users should reconcile this
number in advance. Of course, the question of working Grover's
algorithm with noised components deserves further investigations.
Our results additionally witness that building reliable NISC
devices is very challenging.

\acknowledgments

The authors are grateful to participants of international conference
``Function theory, operator theory and quantum information theory''
(October 4--7 2021, Ufa, Russia), especially to A.S. Holevo and A.S.
Trushechkin, for valuable comments on the subject of this work.

\appendix

\section{On strict positivity of the determinant}\label{app0}

For the given parameter $a$, we define the function
\begin{equation}
f_{a}(\xi)=1-a(\xi+\xi^{2})+\xi^{3}
\, . \label{fadf}
\end{equation}
It is clear for $a\leq0$ that $f_{a}(\xi)>0$ for all $\xi\in[0,1]$.
We claim this inequality for $0<a<1$ as well. For such values of
$a$, one has $f_{a}(0)=1$ and $f_{a}(1)=2(1-a)>0$. After
differentiating with respect to $\xi$, we also obtain
\begin{equation}
\frac{\xdif{f}_{a}(\xi)}{\xdif\xi}=3(\xi-\xi_{+})(\xi-\xi_{-})
\, , \qquad
\xi_{\pm}(a)=\frac{a\pm\sqrt{3a+a^{2}}}{3}
\ . \label{fadf1}
\end{equation}
Due to $\xi_{-}<0$ and $0<\xi_{+}<1$ for $0<a<1$, the critical point
$\xi=\xi_{+}$ is unique in the interval of interest. The function
reaches the minimal value, since around $\xi_{+}$ the derivative
changes from negative to positive. Combining
\begin{equation}
\frac{\xdif}{\xdif{a}}
\>\bigl[\,1-a(\xi_{+}+\xi_{+}^{2})+\xi_{+}^{3}\bigr]
=\!{}-\xi_{+}-\xi_{+}^{2}<0
 \nonumber
\end{equation}
with $f_{1}\bigl(\xi_{+}(1)\bigr)=0$, we show $f_{a}(\xi_{+})>0$ for
$0<a<1$. The latter implies that the function (\ref{fadf}) is
strictly positive for $a<1$ and all $\xi\in[0,1]$. Substituting
$a=\cos2\theta$ and $\xi=\wamma$, we prove strictly positive values
of (\ref{detg}). Indeed, the precondition $M>0$ provides
$\cos2\theta<1$.

\section{Solutions of the recursion equation}\label{app1}

The eigenvalues of (\ref{matra}) are roots of the equation
\begin{equation}
\lambda^{2}-(1+\wamma)\cos2\theta\lambda+\wamma=(\lambda-\lambda_{+})(\lambda-\lambda_{-})=0
\, . \label{chareq1}
\end{equation}
Doing simple algebra leads to
\begin{equation}
2\lambda_{\pm}=(1+\wamma)\cos2\theta\pm\sqrt{(1+\wamma)^{2}\cos^{2}2\theta-4\wamma}
\, . \label{quadeq1}
\end{equation}
The eigenvalues differ whenever
$(1+\wamma)^{2}\cos^{2}2\theta\neq4\wamma$. If so, the matrix $\mtm$
is certainly diagonalizable. Using the corresponding eigenvectors,
we write
\begin{equation}
\mtx^{-1}\mtm\mtx=\dig(\lambda_{+},\lambda_{-})
\, , \label{mtmdg}
\end{equation}
where
\begin{equation}
\mtx=
\begin{pmatrix}
    \wamma\cos2\theta-\lambda_{+} & \wamma\cos2\theta-\lambda_{-} \\
    \sin2\theta & \sin2\theta
\end{pmatrix}
 , \quad
\mtx^{-1}=\frac{1}{\sin2\theta(\lambda_{-}-\lambda_{+})}
\begin{pmatrix}
    \sin2\theta & \lambda_{-}-\wamma\cos2\theta \\
    -\sin2\theta & \wamma\cos2\theta-\lambda_{+}
\end{pmatrix}
 . \nonumber
\end{equation}
If the eigenvalues differ, then $\lambda_{-}-\lambda_{+}\neq0$ with
the existence of $\mtx^{-1}$. To get results for separate point of
the range $\gamma\in[0,1]$, where the eigenvalues are equal, we will
do by continuity. Due to (\ref{vcsv}), one obtains
\begin{equation}
\vcs(t)=\wamma^{t}\,\mtx\,\dig\bigl(\lambda_{+}^{t},\lambda_{-}^{t}\bigr)\,\mtx^{-1}\vcs(0)
\, . \label{sts0}
\end{equation}
Explicitly, we write
\begin{equation}
\begin{pmatrix}
    s_{x}(t)  \\
    s_{z}(t)
\end{pmatrix}
=
\frac{\wamma^{t}}{\sin2\theta(\lambda_{-}-\lambda_{+})}
\begin{pmatrix}
    \lambda_{+}^{t}(\wamma\cos2\theta-\lambda_{+}) & \lambda_{-}^{t}(\wamma\cos2\theta-\lambda_{-}) \\
    \lambda_{+}^{t}\sin2\theta & \lambda_{-}^{t}\sin2\theta
\end{pmatrix}
\begin{pmatrix}
    s_{x}(0)\sin2\theta+s_{z}(0)(\lambda_{-}-\wamma\cos2\theta) \\
    -s_{x}(0)\sin2\theta+s_{z}(0)(\wamma\cos2\theta-\lambda_{+})
\end{pmatrix}
 . \label{sxzt}
\end{equation}
By (\ref{chareq1}) we get $\lambda_{+}\lambda_{-}=\wamma$,
$\lambda_{+}+\lambda_{-}=(1+\wamma)\cos2\theta$, and
$(\wamma\cos2\theta-\lambda_{+})(\wamma\cos2\theta-\lambda_{-})=\wamma\sin^{2}2\theta$.
Up to a factor, the first row of (\ref{sxzt}) is expressed as:
\begin{equation}
\wamma^{-t}s_{x}(t)\sin2\theta(\lambda_{-}-\lambda_{+})
=s_{x}(0)\sin2\theta\bigl[(\lambda_{+}^{t}-\lambda_{-}^{t})\wamma\cos2\theta-\lambda_{+}^{t+1}+\lambda_{-}^{t+1}\bigr]
-s_{z}(0)\wamma\sin^{2}2\theta(\lambda_{+}^{t}-\lambda_{-}^{t})
\, . \label{wamsxt}
\end{equation}
Simplifying expressions, we finally obtain
\begin{equation}
s_{x}(t)=\frac{\wamma^{t}}{\lambda_{-}-\lambda_{+}}
\>\Bigl\{
\wamma(\lambda_{+}^{t}-\lambda_{-}^{t})\bigl[s_{x}(0)\cos2\theta-s_{z}(0)\sin2\theta\bigr]
-(\lambda_{+}^{t+1}-\lambda_{-}^{t+1})s_{x}(0)
\Bigr\}
\, . \label{sxtfin}
\end{equation}
Up to a factor, the second row of (\ref{sxzt}) is similarly expressed as:
\begin{equation}
\wamma^{-t}s_{z}(t)\sin2\theta(\lambda_{-}-\lambda_{+})
=\sin2\theta\,\Bigl\{(\lambda_{+}^{t}-\lambda_{-}^{t})s_{x}(0)\sin2\theta
+(\lambda_{-}^{t}-\lambda_{+}^{t})s_{z}(0)\wamma\cos2\theta
+s_{z}(0)\lambda_{+}\lambda_{-}(\lambda_{+}^{t-1}-\lambda_{-}^{t-1})
\Bigr\}
\, . \label{wamszt}
\end{equation}
Using $\lambda_{+}\lambda_{-}=\wamma$, we finally write
\begin{equation}
s_{z}(t)=\frac{\wamma^{t}}{\lambda_{-}-\lambda_{+}}\>\Bigl\{
(\lambda_{+}^{t}-\lambda_{-}^{t})\bigl[s_{x}(0)\sin2\theta-s_{z}(0)\wamma\cos2\theta\bigr]
+\wamma(\lambda_{+}^{t-1}-\lambda_{-}^{t-1})s_{z}(0)
\Bigr\}
\, . \label{sztfin}
\end{equation}
The Bloch vector after $t$ iterations reads as:
\begin{equation}
\vcr(t)=
\begin{pmatrix}
    s_{x}(t)  \\
    s_{z}(t)
\end{pmatrix}
+
\frac{1-\wamma^{2}}{\det\bigl(\pen_{2}-\wamma\,\mtm\bigr)}
\begin{pmatrix}
    \sin2\theta  \\
    \cos2\theta-\wamma
\end{pmatrix}
 , \label{vcrt}
\end{equation}
where the components of $\vcs(t)$ are given above by (\ref{sxtfin})
and (\ref{sztfin}).

There are two different forms of functional behavior of the Bloch
vector $\vcr(t)$. The first form occurs, when
\begin{equation}
(1+\wamma)^{2}\cos^{2}2\theta<4\wamma
\, , \label{imdi}
\end{equation}
whence $\lambda_{\pm}$ are complex. It holds from
$\lambda_{+}=\lambda_{-}^{*}$ and $\lambda_{+}\lambda_{-}=\wamma$
that
\begin{equation}
\lambda_{\pm}=\wamma^{1/2}\exp(\pm\iu\varphi)
\, , \qquad
\tan\varphi=\frac{\sqrt{4\wamma-(1+\wamma)^{2}\cos^{2}2\theta}}{(1+\wamma)\cos2\theta}
\ . \label{fikin}
\end{equation}
For $\gamma=0$, we obtain $\lambda_{\pm}=\exp(\pm\iu2\theta)$. It
follows from (\ref{fikin}) that
\begin{equation}
\tan^{2}\varphi=\tan^{2}2\theta-
\frac{(1-\wamma)^{2}}{(1+\wamma)^{2}\cos^{2}2\theta}
\ , \label{ph2th}
\end{equation}
whence $\varphi=2\theta+O(\gamma^{2})$ for sufficiently small
$\gamma$. Further, one has
$\lambda_{+}^{t}-\lambda_{-}^{t}=\wamma^{t/2}2\iu\sin{t}\varphi$.
So, the quantities (\ref{sxtfin}) and (\ref{sztfin}) are expressed
via trigonometric functions, namely
\begin{align}
s_{x}(t)&=\wamma^{3t/2}\Bigl\{
\wamma^{1/2}U_{t-1}(\cos\varphi)\bigl[s_{z}(0)\sin2\theta-s_{x}(0)\cos2\theta\bigr]
+U_{t}(\cos\varphi)s_{x}(0)
\Bigr\}
\, , \label{sxtfin1}\\
s_{z}(t)&=\wamma^{3t/2}\Bigl\{
\wamma^{-1/2}U_{t-1}(\cos\varphi)\bigl[s_{z}(0)\wamma\cos2\theta-s_{x}(0)\sin2\theta\bigr]
-U_{t-2}(\cos\varphi)s_{z}(0)
\Bigr\}
\, . \label{sztfin1}
\end{align}
By $U_{t}(\xi)$, we mean here $t$-th Chebyshev polynomial of the
second kind depending on $\xi=\cos\varphi$. These polynomials can be
defined as (see, e.g., \S III.4 of the book \cite{lanczos})
\begin{equation}
U_{t}(\cos\varphi)\sin\varphi=\sin(t+1)\varphi
\qquad (t=0,1,2,\ldots)
\, . \label{chebs}
\end{equation}
It can be seen that $|U_{t}(\xi)|\leq{t}+1$ for $\xi\in[-1,+1]$
and that $U_{t}(\cos\varphi)\approx{t}+1$ for small $t\varphi$. A
linear trend is typical for intermediate zones between peaks and
valleys. A behavior of the terms (\ref{sxtfin1}) and (\ref{sztfin1})
for large $t$ is characterized by product of $\wamma^{3t/2}$ and
some function with at most linear growth.

Another form of functional dependence takes place, when different
eigenvalues are real due to
\begin{equation}
(1+\wamma)^{2}\cos^{2}2\theta>4\wamma
\, . \label{idmi}
\end{equation}
Let us put the parameter $\phi$ such that
$\lambda_{\pm}=\wamma^{1/2}\exp(\pm\phi)$. It follows from
(\ref{quadeq1}) that
\begin{equation}
\cosh\phi=\frac{(1+\wamma)\cos2\theta}{2\wamma^{1/2}}
\ . \label{shph}
\end{equation}
Then we have
$\lambda_{+}^{t}-\lambda_{-}^{t}=\wamma^{t/2}2\sinh{t}\phi$. The
components (\ref{sxtfin}) and (\ref{sztfin}) can be rewritten in the
form:
\begin{align}
s_{x}(t)&=\wamma^{3t/2}\Bigl\{
\wamma^{1/2}U_{t-1}(\cosh\phi)\bigl[s_{z}(0)\sin2\theta-s_{x}(0)\cos2\theta\bigr]
+U_{t}(\cosh\phi)s_{x}(0)
\Bigr\}
\, , \label{sxtfin11}\\
s_{z}(t)&=\wamma^{3t/2}\Bigl\{
\wamma^{-1/2}U_{t-1}(\cosh\phi)\bigl[s_{z}(0)\wamma\cos2\theta-s_{x}(0)\sin2\theta\bigr]
-U_{t-2}(\cosh\phi)s_{z}(0)
\Bigr\}
\, . \label{sztfin11}
\end{align}
For Chebyshev polynomials of the second kind, we adopt here the
expression
\begin{equation}
U_{t}(\xi)=
\frac{\bigl(\xi+\sqrt{\xi^{2}-1}\,\bigr)^{t+1}-\bigl(\xi-\sqrt{\xi^{2}-1}\,\bigr)^{t+1}}{2\sqrt{\xi^{2}-1}}
\ . \label{chebs11}
\end{equation}
In contrast to (\ref{sxtfin1}) and (\ref{sztfin1}), we now deal with
functions of the form $U_{t}(\cosh\phi)$. They are hyperbolic and
rise exponentially with $t$. However, if the parameters are such
that $t\phi\ll1$, then the corresponding terms again have a linear
growth with increasing $t$. This shows a continuity of picture, when
$\gamma$ is increased so that (\ref{imdi}) is replaced with
(\ref{idmi}). Of course, the claim about linear growth is applicable
only for limited ranges of iterations. As was already mentioned, it
is advisable to use only iterations sufficient to get a vicinity of
the first peak.

\end{document}